\documentclass[12pt]{article}
\usepackage{amssymb,amsthm}

\title{Gaudin Hypothesis for the $XYZ$ Spin Chain}
\author{Yasuhiro Fujii\footnotemark[2]\ \ and Miki Wadati}
\date{\textit{\small Department of Physics, Graduate School of Science,
  University of Tokyo,\\  Hongo 7--3--1, Bunkyo-ku, Tokyo 113--0033, Japan}}

\renewcommand{\theequation}{\arabic{section}.\arabic{equation}}

\newcommand{\rmi}{\mbox{i}}
\newcommand{\rme}{\mbox{e}}
\newcommand{\rmd}{\mbox{d}}
\newcommand{\rmii}{\mbox{\scriptsize i}}
\newcommand{\der}{\partial}
\newcommand{\M}{\mathcal{M}}
\renewcommand{\|}{|\!|}
\renewcommand{\epsilon}{\varepsilon}
\renewcommand{\phi}{\varphi}

\theoremstyle{plain}
\newtheorem*{theorem}{Theorem}
\theoremstyle{definition}
\newtheorem{lemma}{Lemma}

\begin{document}

\maketitle
\footnotetext[2]{e-mail: \texttt{fujii@monet.phys.s.u-tokyo.ac.jp}}

\begin{abstract}
  The $XYZ$ spin chain is considered
  in the framework of the generalized algebraic Bethe ansatz
  developed by Takhtajan and Faddeev.
  The sum of norms of the Bethe vectors is computed
  and expressed in the form of a Jacobian.
  This result corresponds to the Gaudin hypothesis
  for the $XYZ$ spin chain.
\end{abstract}

\section{Introduction}

In this paper we consider the $XYZ$ spin chain
with the periodic boundary condition.
The Hamiltonian is defined by
\begin{equation}
  \label{h}
  H_{XYZ} = -\frac{1}{2}\sum_{n=1}^L
  (J_x\sigma_n^x\sigma_{n+1}^x+J_y\sigma_n^y\sigma_{n+1}^y
  +J_z\sigma_n^z\sigma_{n+1}^z).
\end{equation}
Here $\sigma_n^x$, $\sigma_n^y$ and $\sigma_n^z$
are the Pauli matrices acting on a Hilbert state space $H_n=\mathbb{C}^2$.
The Hamiltonian thus acts on $\otimes_{n=1}^L H_n$.
The coupling constants
$J_x$, $J_y$ and $J_z$ are parameterized by
\begin{equation}
  J_x = 1+k\mbox{sn}^2 2\eta,
  \qquad
  J_y = 1-k\mbox{sn}^2 2\eta,
  \qquad
  J_z = \mbox{cn}2\eta\mbox{dn}2\eta,
\end{equation}
where $k$ is the modulus of the Jacobi elliptic functions.
In the limit $k\rightarrow 0$ $J_x$, $J_y$ and $J_z$ satisfy
$J_x=J_y=1$ and $J_z=\cos 2\eta$,
and the $XYZ$ spin chain is reduced to the $XXZ$ spin chain.

The $XYZ$ spin chain was first solved
by Baxter in a series of remarkable papers \cite{B1,B2}.
He discovered a link between the $XYZ$ spin chain
and a two-dimensional classical model, the so-called eight-vertex model,
and obtained a system of transcendental equations.
With the help of these equations
the energy of the ground state of the $XYZ$ spin chain was calculated.
Furthermore, he found the eigenvectors and eigenvalues
of the $XYZ$ spin chain
by means of a generalization of the Bethe ansatz method \cite{B3}.
Referring to the algebraic Bethe ansatz,
which is more intelligible than the Bethe ansatz,
Takhtajan and Faddeev succeeded in
simplifying this Baxter's method \cite{TF}.
Their method is called
the \textit{generalized algebraic Bethe ansatz}
and enables us to deal with the $XYZ$ spin chain more systematically.

For the $XXZ$ spin chain,
by means of the usual algebraic Bethe ansatz,
various correlation functions have been calculated.
Gaudin forecasted that norms of the eigenvectors
are expressed by Jacobians,
and Korepin proved his hypothesis \cite{K}. 
Based on this fact
scalar products of arbitrary vectors were shown to be
represented by determinants of matrices
that contain bosonic quantum fields called the dual fields \cite{EFIK}.
Using them
one can evaluate any correlation function
of the $XXZ$ spin chain.
Recently, these results have been extended to the asymmetric $XXZ$ chain
that is a non-hermitian generalization of the $XXZ$ spin chain \cite{FW}.

The aim of the paper is to prove the Gaudin hypothesis
for the $XYZ$ spin chain by using the generalized algebraic Bethe ansatz.
We show that the sum of norms of the Bethe vectors
is expressed by a Jacobian.
However, norms of the eigenvectors
can not be computed in the framework
of the generalized algebraic Bethe ansatz,
because the Bethe vectors are not equivalent to the eigenvectors
(see (\ref{y})--(\ref{dy}) and (\ref{f})--(\ref{df})).
We interpret the Gaudin hypothesis as a theorem
that holds for the Bethe vectors.
This interpretation is supported by the fact that
the Bethe vectors correspond to the eigenvectors
in the usual algebraic Bethe ansatz for the $XXZ$ spin chain.

Our result lays the foundation of calculation of correlation functions
of the $XYZ$ spin chain.
In section~\ref{gba} we review
the generalized algebraic Bethe ansatz.
In section~\ref{g} the sum of norms of the Bethe vectors
is shown to be given by a Jacobian.
Section~\ref{con} is devoted to concluding remarks.

\setcounter{equation}{0}
\renewcommand{\theequation}
{\arabic{section}.\arabic{subsection}.\arabic{equation}}

\section{Generalized Algebraic Bethe Ansatz}
\label{gba}

In this section we review the generalized algebraic Bethe ansatz
for the $XYZ$ spin chain.
In the original paper \cite{TF}
the dual eigenvectors are not investigated.
We include them for the first time.

\subsection{Description of the model}
Central objects of the generalized algebraic Bethe ansatz
are the $R$-matrix and the $L$-operator.
The $R$-matrix is of the form:
\begin{equation}
  R(\lambda,\mu) =
  \left(
    \begin{array}{cccc}
      a(\lambda,\mu) & 0 & 0 & d(\lambda,\mu) \\
      0 & b(\lambda,\mu) & c(\lambda,\mu) & 0 \\
      0 & c(\lambda,\mu) & b(\lambda,\mu) & 0 \\
      d(\lambda,\mu) & 0 & 0 & a(\lambda,\mu)
    \end{array}
  \right).
\end{equation}
where
\begin{equation}
  \begin{array}{lll}
    a(\lambda,\mu) &=&
    \Theta(2\eta)\Theta(\lambda-\mu)H(\lambda-\mu+2\eta), \\
    b(\lambda,\mu) &=&
    H(2\eta)\Theta(\lambda-\mu)\Theta(\lambda-\mu+2\eta), \\
    c(\lambda,\mu) &=&
    \Theta(2\eta)H(\lambda-\mu)\Theta(\lambda-\mu+2\eta), \\
    d(\lambda,\mu) &=&
    H(2\eta)H(\lambda-\mu)H(\lambda-\mu+2\eta).
  \end{array}
\end{equation}
We call $\lambda,\mu\in\mathbb{C}$ the spectral parameters.
$H(\mu)$ and $\Theta(\mu)$ are the Jacobi theta functions
with quasi-periods $2K,2\rmi K'\in\mathbb{C}$
(Im $\rmi K'/K>0$).
In this paper we assume that there exist $Q\in\mathbb{Z}_{>0}$
such that
\begin{equation}
  \label{qm}
  Q\eta = 2K.
\end{equation}
Then $H(\mu)$ and $\Theta(\mu)$ have a period $2Q\eta$:
\begin{equation}
  H(\mu+2Q\eta) = H(\mu),
  \qquad
  \Theta(\mu+2Q\eta) = \Theta(\mu).
\end{equation}
The $L$-operator is expressed by a $2\times 2$ matrix
whose elements contain the Pauli matrices:
\begin{equation}
  L_n(\mu) =
  \left(
    \begin{array}{cc}
      w_4+w_3\sigma_n^z & w_1\sigma_n^x-\rmi w_2\sigma_n^y \\
      w_1\sigma_n^x+\rmi w_2\sigma_n^y & w_4-w_3\sigma_n^z
    \end{array}
  \right),
\end{equation}
where
\begin{equation}
  \begin{array}{lll}
    w_4+w_3 &=& \Theta(2\eta)\Theta(\mu-\eta)H(\mu+\eta), \\
    w_4-w_3 &=& \Theta(2\eta)H(\mu-\eta)\Theta(\mu+\eta), \\
    w_1+w_2 &=& H(2\eta)\Theta(\mu-\eta)\Theta(\mu+\eta), \\
    w_1-w_2 &=& H(2\eta)H(\mu-\eta)H(\mu+\eta).
  \end{array}
\end{equation}
The $L$-operator $L_n(\mu)$ acts on a Hilbert state space $H_n$.
This satisfies the \textit{Yang--Baxter equation}:
\begin{equation}
  \label{preybe}
  R(\lambda,\mu)(L_n(\lambda)\otimes L_n(\mu)) =
  (L_n(\mu)\otimes L_n(\lambda))R(\lambda,\mu).
\end{equation}

The Hamiltonian is derived from the $L$-operator as follows.
The product of the $L$-operators is called the monodromy matrix
and is expressed in a $2\times 2$ matrix form:
\begin{equation}
  T(\mu) =
  \prod_{n=1}^{\longleftarrow \atop{L}}L_n(\mu) =
  \left(
    \begin{array}{cc}
      A(\mu) & B(\mu) \\
      C(\mu) & D(\mu)
    \end{array}
  \right).
\end{equation}
The trace of the monodromy matrix over matrix space
\begin{equation}
  t(\mu) = \mbox{tr}T(\mu) = A(\mu)+D(\mu),
\end{equation}
is called the transfer matrix
and gives the Hamiltonian (\ref{h}) via
\begin{equation}
  H_{XYZ} =
  -\mbox{sn}2\eta\left.\frac{\rmd}{\rmd\mu}\log t(\mu)\right|_{\mu=\eta}
  +\mbox{const.}
\end{equation}
The Hamiltonian is thus diagonalized by the eigenvectors
of the transfer matrix.

\setcounter{equation}{0}
\subsection{Gauge transformations}
We introduce a family of gauge transformations
with free parameters $s,t\in\mathbb{C}$
and integer $l=0,\ldots,Q-1$.
The $L$-operator is replaced by
\begin{equation}
  L_n^l(\mu) = M_{n+l}^{-1}(\mu)L_n(\mu)M_{n+l-1}(\mu) =
  \left(
    \begin{array}{cc}
      \alpha_n^l(\mu) & \beta_n^l(\mu) \\
      \gamma_n^l(\mu) & \delta_n^l(\mu)
    \end{array}
  \right),
\end{equation}
with matrices $M_k(\mu)$ ($k=0,\ldots,Q-1$) defined by
\begin{equation}
  M_k(\mu) =
  \left(
    \begin{array}{cc}
      H(s+2k\eta-\mu) & (g(\tau_k))^{-1}H(t+2k\eta+\mu) \\
      \Theta(s+2k\eta-\mu) & (g(\tau_k))^{-1}\Theta(t+2k\eta+\mu)
    \end{array}
  \right),
\end{equation}
where
\begin{equation}
  \tau_k = \frac{s+t}{2}+2k\eta-K,
  \qquad
  g(\mu) = H(\mu)\Theta(\mu).
\end{equation}
In the generalized algebraic Bethe ansatz
the following vectors are important:
\begin{eqnarray}
  |\omega_n^l\rangle &=&
  H(s+(2(n+l)-1)\eta)|\uparrow\rangle_n
  +\Theta(s+(2(n+l)-1)\eta)|\downarrow\rangle_n, \\
  \langle\omega_n^l| &=&
  {}_n\langle\uparrow|\Theta(t+(2(n+l)-1)\eta)
  -{}_n\langle\uparrow|H(t+(2(n+l)-1)\eta).
\end{eqnarray}
Here $|\uparrow\rangle_n$ and $|\downarrow\rangle_n$
are the orthonormal basis of a Hilbert state space $H_n$,
and ${}_n\langle\uparrow|$ and ${}_n\langle\downarrow|$
are those dual basis.
The actions of elements of the transformed $L$-operator
on $|\omega_n^l\rangle$ and $\langle\omega_n^l|$ are computed as follows:
\begin{eqnarray}
  \label{aw}
  \alpha_n^l(\mu)|\omega_n^l\rangle &=&
  h(\mu+\eta)|\omega_n^{l-1}\rangle, \\
  \delta_n^l(\mu)|\omega_n^l\rangle &=&
  h(\mu-\eta)|\omega_n^{l+1}\rangle, \\
  \gamma_n^l(\mu)|\omega_n^l\rangle &=& 0, \\
  \langle\omega_n^l|\alpha_n^l(\mu) &=&
  \langle\omega_n^{l+1}|\frac{g(\tau_{n+l-1})}{g(\tau_{n+l})}h(\mu+\eta), \\
  \langle\omega_n^l|\delta_n^l(\mu) &=&
  \langle\omega_n^{l-1}|\frac{g(\tau_{n+l})}{g(\tau_{n+l-1})}h(\mu-\eta), \\
  \label{wb}
  \langle\omega_n^l|\beta_n^l(\mu) &=& 0,
\end{eqnarray}
where $h(\mu)=g(\mu)\Theta(0)$.
Note that $|\omega_n^l\rangle$ and $\langle\omega_n^l|$
are independent of the spectral parameters.
They are called the local vacuums.

For $k,l=0,\ldots,Q-1$ we introduce a matrix
\begin{equation}
  T_{k,l}(\mu) =
  M_k^{-1}(\mu)T(\mu)M_l(\mu) =
  \left(
    \begin{array}{cc}
      A_{k,l}(\mu) & B_{k,l}(\mu) \\
      C_{k,l}(\mu) & D_{k,l}(\mu)
    \end{array}
  \right).
\end{equation}
Under the gauge transformations
the monodromy matrix $T(\mu)$ is replaced by $T_{L+l,l}(\mu)$.
We thus call $T_{k,l}(\mu)$ the generalized monodromy matrix.
This plays a central role in the next sub-section.

The products of the local vacuums are
called the generating vectors:
\begin{equation}
  |l\rangle = |\omega_L^l\rangle\otimes\cdots\otimes|\omega_1^l\rangle,
  \qquad
  \langle l| = \langle\omega_1^l|\otimes\cdots\otimes\langle\omega_L^l|.
\end{equation}
By use of local formulae (\ref{aw})--(\ref{wb})
the actions of elements of the monodromy matrix
on the generating vectors are computed as follows:
\begin{eqnarray}
  \label{al}
  A_{L+l,l}(\mu)|l\rangle &=& (h(\mu+\eta))^L|l-1\rangle, \\
  D_{L+l,l}(\mu)|l\rangle &=& (h(\mu-\eta))^L|l+1\rangle, \\
  C_{L+l,l}(\mu)|l\rangle &=& 0, \\
  \langle l|A_{L+l,l}(\mu) &=&
  \langle l+1|\frac{g(\tau_l)}{g(\tau_{L+l})}(h(\mu+\eta))^L, \\
  \langle l|D_{L+l,l}(\mu) &=&
  \langle l-1|\frac{g(\tau_{L+l})}{g(\tau_l)}(h(\mu-\eta))^L, \\
  \label{lb}
  \langle l|B_{L+l,l}(\mu) &=& 0.
\end{eqnarray}
If $Q$ divides $L$
extra factors of $g(\tau_l)$ and $g(\tau_{L+l})$ are canceled.
Hereafter we assume that
the lattice length $L$ is multiple of $Q$.

\setcounter{equation}{0}
\subsection{Generalized algebraic Bethe ansatz}
The Yang--Baxter equation (\ref{preybe}) can be shifted up to
the level of the monodromy matrix:
\begin{equation}
  \label{ybe}
  R(\lambda,\mu)(T(\lambda)\otimes T(\mu)) =
  (T(\mu)\otimes T(\lambda))R(\lambda,\mu).
\end{equation}
From this Yang--Baxter equation
one can obtain the commutation relations among elements
of the generalized monodromy matrix.
Useful relations are the following:
\begin{eqnarray}
  A_{k,l}(\lambda)A_{k+1,l+1}(\mu) &=&
  A_{k,l}(\mu)A_{k+1,l+1}(\lambda), \\
  \label{bb}
  B_{k,l+1}(\lambda)B_{k+1,l}(\mu) &=&
  B_{k,l+1}(\mu)B_{k+1,l}(\lambda), \\
  \label{cc}
  C_{k+1,l}(\lambda)C_{k,l+1}(\mu) &=&
  C_{k+1,l}(\mu)C_{k,l+1}(\lambda), \\
  D_{k+1,l+1}(\lambda)D_{k,l}(\mu) &=&
  D_{k+1,l+1}(\mu)D_{k,l}(\lambda),
\end{eqnarray}
\begin{eqnarray}
  \label{ab}
  A_{k,l}(\lambda)B_{k+1,l-1}(\mu) &=&
  \alpha(\lambda,\mu)B_{k,l-2}(\mu)A_{k+1,l-1}(\lambda)
  \nonumber \\ && \qquad
  -\beta_{l-1}(\lambda,\mu)B_{k,l-2}(\lambda)A_{k+1,l-1}(\mu), \\
  D_{k,l}(\lambda)B_{k+1,l-1}(\mu) &=&
  \alpha(\mu,\lambda)B_{k+2,l}(\mu)D_{k+1,l-1}(\lambda)
  \nonumber \\ && \qquad
  +\beta_{k+1}(\lambda,\mu)B_{k+2,l}(\mu)D_{k+1,l-1}(\lambda), \\
  C_{k-1,l-1}(\mu)A_{k,l}(\lambda) &=&
  \alpha(\lambda,\mu)A_{k+1,l-1}(\lambda)C_{k,l}(\mu)
  \nonumber \\ && \qquad
  +\beta_k(\mu,\lambda)A_{k+1,l-1}(\mu)C_{k,l}(\lambda), \\
  \label{cd}
  C_{k+1,l+1}(\mu)D_{k,l}(\lambda) &=&
  \alpha(\mu,\lambda)D_{k+1,l-1}(\lambda)C_{k,l}(\mu)
  \nonumber \\ && \qquad
  -\beta_l(\mu,\lambda)D_{k+1,l-1}(\mu)C_{k,l}(\lambda),
\end{eqnarray}
\begin{eqnarray}
  \label{cb}
  \lefteqn{C_{k-1,l+1}(\lambda)B_{k,l}(\mu)
    -\frac{g(\tau_{l-1})g(\tau_{l+1})}{g^2(\tau_l)}
    B_{k+1,l-1}(\mu)C_{k,l}(\lambda)}
  \nonumber \\ &=&
  \beta_k(\lambda,\mu)A_{k+1,l+1}(\lambda)D_{k,l}(\mu)
  -\beta_l(\lambda,\mu)A_{k+1,l+1}(\mu)D_{k,l}(\lambda),
\end{eqnarray}
where
\begin{equation}
  \alpha(\lambda,\mu) =
  \frac{h(\lambda-\mu-2\eta)}{h(\lambda-\mu)},
  \qquad
  \beta_k(\lambda,\mu) =
  \frac{h(2\eta)}{h(\mu-\lambda)}
  \frac{h(\tau_k+\mu-\lambda)}{h(\tau_k)}.
\end{equation}

The generalized algebraic Bethe ansatz offers a simple method
to find the eigenvectors and eigenvalues of the transfer matrix:
\begin{equation}
  t(\mu) = \mbox{tr}T(\mu) = A_{l,l}(\mu)+D_{l,l}(\mu).
\end{equation}
Let us introduce vectors
\begin{eqnarray}
  \label{y}
  |\Psi_l(\lambda_1,\ldots,\lambda_N)\rangle &=&
  B_{l+1,l-1}(\lambda_1)\cdots B_{l+N,l-N}(\lambda_N)|l-N\rangle, \\
  \label{dy}
  \langle\Psi_l(\lambda_1,\ldots,\lambda_N)| &=&
  \langle l-N+1|C_{l+N-1,l-N+1}(\lambda_N)\cdots C_{l,l}(\lambda_1).
\end{eqnarray}
Here we set
\begin{equation}
  \label{n}
  2N \equiv 0 \bmod Q.
\end{equation}
Namely, the admissible values of $N$ are
\[
  N = \left\{
    \begin{array}{ll}
      0,Q,2Q,\ldots,L & \mbox{for odd } Q \\
      0,Q/2,Q,\ldots,L & \mbox{for even } Q.
    \end{array}
  \right.
\]
Referring to the algebraic Bethe ansatz for the $XXZ$ spin chain
we call the vectors (\ref{y})--(\ref{dy}) the Bethe vectors.
By means of commutation relations (\ref{bb})--(\ref{cc}),
(\ref{ab})--(\ref{cd})
and relations (\ref{al})--(\ref{lb})
the actions of $A_{l,l}(\mu)$ and $D_{l,l}(\mu)$
on the Bethe vectors are computed as follows:
\begin{eqnarray}
  \label{ay}
  A_{l,l}(\mu)|\Psi_l(\lambda_1,\ldots,\lambda_N)\rangle &=&
  {}_1\Lambda(\mu;\{\lambda_k\})
  |\Psi_{l-1}(\lambda_1,\ldots,\lambda_N)\rangle
  \nonumber \\ && \hspace{-80pt}
  +\sum_{j=1}^N{}_1\Lambda_j^l(\mu;\{\lambda_k\})
  |\Psi_{l-1}(\lambda_1,\ldots,\lambda_{j-1},
  \mu,\lambda_{j+1},\ldots,\lambda_N)\rangle, \\
  \label{ddy}
  D_{l,l}(\mu)|\Psi_l(\lambda_1,\ldots,\lambda_N)\rangle &=&
  {}_2\Lambda(\mu;\{\lambda_k\})
  |\Psi_{l+1}(\lambda_1,\ldots,\lambda_N)\rangle
  \nonumber \\ && \hspace{-80pt}
  +\sum_{j=1}^N{}_2\Lambda_j^l(\mu;\{\lambda_k\})
  |\Psi_{l+1}(\lambda_1,\ldots,\lambda_{j-1},
  \mu,\lambda_{j+1},\ldots,\lambda_N)\rangle, \\
  \langle\Psi_{l-1}(\lambda_1,\ldots,\lambda_N)|A_{l,l}(\mu) &=&
  \langle\Psi_l(\lambda_1,\ldots,\lambda_N)|
  {}_1\Lambda(\mu;\{\lambda_k\})
  \nonumber \\ && \hspace{-80pt}
  +\sum_{j=1}^N\langle\Psi_l(\lambda_1,\ldots,\lambda_{j-1},
  \mu,\lambda_{j+1},\ldots,\lambda_N)|
  {}_1\widetilde{\Lambda}_j^l(\mu;\{\lambda_k\}), \\
  \label{yd}
  \langle\Psi_{l+1}(\lambda_1,\ldots,\lambda_N)|D_{l,l}(\mu) &=&
  \langle\Psi_l(\lambda_1,\ldots,\lambda_N)|
  {}_2\Lambda(\mu;\{\lambda_k\})
  \nonumber \\ && \hspace{-80pt}
  +\sum_{j=1}^N\langle\Psi_l(\lambda_1,\ldots,\lambda_{j-1},
  \mu,\lambda_{j+1},\ldots,\lambda_N)|
  {}_2\widetilde{\Lambda}_j^l(\mu;\{\lambda_k\}),  
\end{eqnarray}
where
\begin{eqnarray}
  {}_1\Lambda(\mu;\{\lambda_k\}) &=&
  (h(\mu+\eta))^L\prod_{k=1}^N\alpha(\mu,\lambda_k), \\
  {}_2\Lambda(\mu;\{\lambda_k\}) &=&
  (h(\mu-\eta))^L\prod_{k=1}^N\alpha(\lambda_k,\mu), \\
  {}_1\Lambda_j^l(\mu;\{\lambda_k\}) &=&
  -\beta_{l-1}(\mu,\lambda_j)(h(\lambda_j+\eta))^L
  \prod_{k\ne j}^N\alpha(\lambda_j,\lambda_k), \\
  {}_2\Lambda_j^l(\mu;\{\lambda_k\}) &=&
  \beta_{l+1}(\mu,\lambda_j)(h(\lambda_j-\eta))^L
  \prod_{k\ne j}^N\alpha(\lambda_k,\lambda_j), \\
  {}_1\widetilde{\Lambda}_j^l(\mu;\{\lambda_k\}) &=&
  \beta_l(\lambda_j,\mu)(h(\lambda_j+\eta))^L
  \prod_{k\ne j}^N\alpha(\lambda_j,\lambda_k), \\
  {}_2\widetilde{\Lambda}_j^l(\mu;\{\lambda_k\}) &=&
  -\beta_l(\lambda_j,\mu)(h(\lambda_j-\eta))^L
  \prod_{k\ne j}^N\alpha(\lambda_k,\lambda_j).
\end{eqnarray}
For integer $m=0,\ldots,Q-1$
consider the following linear combinations of the Bethe vectors:
\begin{eqnarray}
  \label{f}
  |\Phi_m(\lambda_1,\ldots,\lambda_N)\rangle &=&
  \frac{1}{\sqrt{Q}}\sum_{l=0}^{Q-1}\rme^{2\pi\rmii lm/Q}
  |\Psi_l(\lambda_1,\ldots,\lambda_N)\rangle, \\
  \label{df}
  \langle\Phi_m(\lambda_1,\ldots,\lambda_N)| &=&
  \frac{1}{\sqrt{Q}}\sum_{l=0}^{Q-1}
  \langle\Psi_l(\lambda_1,\ldots,\lambda_N)|\rme^{-2\pi\rmii lm/Q}.
\end{eqnarray}
By means of relations (\ref{ay})--(\ref{yd})
they are shown to be the eigenvectors of the transfer matrix:
\begin{eqnarray}
  t(\mu)|\Phi_m(\lambda_1,\ldots,\lambda_N)\rangle &=&
  \Lambda_m(\mu;\{\lambda_k\})|\Phi_m(\lambda_1,\ldots,\lambda_N)\rangle, \\
  \langle\Phi_m(\lambda_1,\ldots,\lambda_N)|t(\mu) &=&
  \langle\Phi_m(\lambda_1,\ldots,\lambda_N)|\Lambda_m(\mu;\{\lambda_k\}),
\end{eqnarray}
if the spectral parameters $\{\lambda_j\}$
satisfy the \textit{Bethe ansatz equations}:
\begin{equation}
  \label{prebe}
  \left(\frac{h(\lambda_j+\eta)}{h(\lambda_j-\eta)}\right)^L =
  \rme^{-4\pi\rmii m/Q}\prod_{k\ne j}^N
  \frac{\alpha(\lambda_k,\lambda_j)}{\alpha(\lambda_j,\lambda_k)}.
  \qquad
  (j=1,\ldots,N)
\end{equation}
Here the eigenvalue is given by
\begin{equation}
  \Lambda_m(\mu;\{\lambda_k\}) =
  \rme^{2\pi\rmii m/Q}{}_1\Lambda(\mu;\{\lambda_k\})
  +\rme^{-2\pi\rmii m/Q}{}_2\Lambda(\mu;\{\lambda_k\}).
\end{equation}
We thus have obtained the eigenvectors for the $XYZ$ spin chain
(\ref{f})--(\ref{df}).

In the case $Q=2$
the Bethe ansatz equations break up into
$N$ independent equations for the spectral parameters $\{\lambda_j\}$.
This case corresponds to the Ising, dimer and
free-fermion models \cite{B1}.

\setcounter{equation}{0}
\renewcommand{\theequation}{\arabic{section}.\arabic{equation}}
\section{Gaudin Hypothesis}
\label{g}

In this section we compute the sum of norms of the Bethe vectors:
\begin{equation}
  \M_n(\lambda_1,\ldots,\lambda_n) =
  \frac{1}{Q}\sum_{l=0}^{Q-1}
  \langle\Psi_l^n(\lambda_1,\ldots,\lambda_n)|
  \Psi_l^n(\lambda_1,\ldots,\lambda_n)\rangle.
\end{equation}
Here the Bethe vectors are redefined by
\begin{eqnarray}
  |\Psi_l^n(\lambda_1,\ldots,\lambda_n)\rangle &=&
  B_{l+N-n+1,l-N+n-1}(\lambda_1)\cdots B_{l+N,l-N}(\lambda_n)|l-N\rangle, \\
  \langle\Psi_l^n(\lambda_1,\ldots,\lambda_n)| &=&
  \langle l-N+1|C_{l+N-1,l-N+1}(\lambda_n)\cdots C_{l+N-n,l-N+n}(\lambda_1),
\end{eqnarray}
and the spectral parameters $\{\lambda_j\}$ are supposed
to satisfy the Bethe ansatz equations:
\begin{equation}
  \label{be}
  r(\lambda_j)\prod_{k\ne j}^n
  \frac{\alpha(\lambda_j,\lambda_k)}{\alpha(\lambda_k,\lambda_j)} =
  \rme^{-4\pi\rmii m/Q},
  \qquad
  (j=1,\ldots,n)
\end{equation}
where
\begin{equation}
  r(\lambda) =
  \left(\frac{h(\lambda+\eta)}{h(\lambda-\eta)}\right)^L.
\end{equation}
We compute $\M_n(\lambda_1,\ldots,\lambda_n)$ by induction on $n$.
Let
\begin{equation}
  \|\lambda_1,\ldots,\lambda_n\|_n =
  \frac{(-h'(0))^n \M_n(\lambda_1,\ldots,\lambda_n)}
  {\displaystyle
    c_L(h(2\eta))^n\prod_{j=1}^n(h(\lambda_j+\eta)h(\lambda_j-\eta))^L
    \prod_{j\ne k}^n\alpha(\lambda_j,\lambda_k)},
\end{equation}
with the norm of the generating vectors:
\begin{equation}
  c_L =
  \langle l|l-1\rangle =
  \left(\frac{2g(\eta-\frac{s-t}{2})}{g(K)}\right)^L
  \prod_{i=1}^L g(\tau_{i+l-2}).
\end{equation}
Notice that $c_L$
is independent of $l$ due to the periodicity of $g(\mu)$.

Extending Korepin's proof of the Gaudin hypothesis \cite{K}
we prove that $\|\lambda_1,\ldots,\lambda_n\|_n$ is expressed
in the form of a Jacobian (see (\ref{result})).
This result implies the Gaudin hypothesis for the $XYZ$ spin chain;
the Gaudin hypothesis is regarded as a theorem
that holds for the Bethe vectors
by virtue of the fact that they correspond to the eigenvectors
in the usual algebraic Bethe ansatz.

Using the solutions of the Bethe ansatz equations $\{\lambda_k\}$
we introduce new parameters:
\begin{equation}
  X_j = \frac{\rmd}{\rmd\lambda_j}\log r(\lambda_j).
  \qquad
  (j=1,\ldots,n)
\end{equation}
\begin{lemma}
  \label{l1}
  $\|\lambda_1,\ldots,\lambda_n\|_n$ is invariant
  under simultaneous replacements:
  \[
    \lambda_j\leftrightarrow\lambda_k 
    \quad\mbox{and}\quad
    X_j\leftrightarrow X_k.
    \qquad
    (j,k=1,\ldots,n)
  \]
\end{lemma}
\begin{proof}
  Because of commutation relations (\ref{bb})--(\ref{cc}),
  $\M_n(\lambda_1,\ldots,\lambda_n)$ and therefore
  $\|\lambda_1,\ldots,\lambda_n\|_n$ are invariant
  under the replacements.
\end{proof}
\begin{lemma}
  \label{l2}
  $\|\lambda_1,\ldots,\lambda_n\|_n=0$ if $X_1=\cdots=X_n=0$.
\end{lemma}
\begin{proof}
  Let $4\epsilon=\min_{j\ne k}|\lambda_j-\lambda_k|$
  and consider a new continuous function
  $\tilde{r}(\lambda)$ that coincides with $r(\lambda_j)$
  for $|\lambda-\lambda_j|\leqslant\epsilon$ ($j,k=1,\ldots,n$).
  By definition
  the set $\{X_j\}$ derived from $\tilde{r}(\lambda)$
  satisfies $X_1=\cdots=X_n=0$.
  Next, we introduce new spectral parameters
  \begin{equation}
    \tilde{\lambda}_j = \lambda_j+\delta,
    \qquad
    |\delta|<\epsilon.
    \qquad
    (j=1,\ldots,n)
  \end{equation}
  These spectral parameters $\{\tilde{\lambda}_j\}$
  still obey the Bethe ansatz equations (\ref{be}),
  because $\alpha(\tilde{\lambda}_j,\tilde{\lambda}_k)$ depends only on
  $\lambda_j-\lambda_k$
  and $\tilde{r}(\tilde{\lambda}_j)$ is equal to $r(\lambda_j)$
  by definitions of $\tilde{r}(\lambda)$ and $\{\tilde{\lambda}_j\}$
  ($j,k=1,\ldots,n$).
  We define
  \begin{equation}
    F_n(\delta) =
    \frac{1}{Q}\sum_{l=0}^{Q-1}
    \langle\Psi_l^n(\lambda_1,\ldots,\lambda_n)|
    \Psi_l^n(\tilde{\lambda}_1,\ldots,\tilde{\lambda}_n)\rangle.
  \end{equation}
  Evaluating $F_n(\delta)$ helps us to prove the lemma.
  Compute
  \begin{eqnarray*}
    \frac{1}{Q}\sum_{l=0}^{Q-1}
    (\rme^{2\pi\rmii m/Q}\langle\Psi_{l-1}^n(\lambda_1,\ldots,\lambda_n)|
    A_{l+N-n,l-N+n}(\mu)
    |\Psi_l^n(\tilde{\lambda}_1,\ldots,\tilde{\lambda}_n)\rangle
    \nonumber \\
    +\rme^{-2\pi\rmii m/Q}\langle\Psi_{l+1}^n(\lambda_1,\ldots,\lambda_n)|
    D_{l+N-n,l-N+n}(\mu)
    |\Psi_l^n(\tilde{\lambda}_1,\ldots,\tilde{\lambda}_n)\rangle)
  \end{eqnarray*}
  in two ways that
  both of $A_{l+N-n,l-N+n}(\mu)$ and $D_{l+N-n,l-N+n}(\mu)$
  operate to the left or to the right.
  It thus follows that
  \begin{equation}
    (\Lambda_m(\mu;\{\lambda_k\})-\Lambda_m(\mu;\{\tilde{\lambda}_k\}))
    F_n(\delta) = 0.
  \end{equation}
  Since $\Lambda_m(\mu;\{\lambda_k\})$ is a continuous function
  for $\{\lambda_k\}$, $F_n(\delta)$ must be $0$.
  Due to the definition of $\|\lambda_1,\ldots,\lambda_n\|_n$
  the proof has been completed.
\end{proof}
\begin{lemma}
  \label{l3}
  $\|\lambda_1,\ldots,\lambda_n\|_n$ satisfies a recursion relation:
  \begin{equation}
    \|\lambda_1,\ldots,\lambda_n\|_n =
    \|\lambda_2,\ldots,\lambda_n\|_{n-1}^{\mbox{\scriptsize mod}}X_1+V_1,
  \end{equation}
  where $V_1$ is independent of $X_1$.
  $\|\lambda_2,\ldots,\lambda_n\|_{n-1}^{\mbox{\scriptsize mod}}$
  is defined by $n-1$ solutions of the Bethe ansatz equations
  and $r(\lambda)$ is modified by
  \begin{equation}
    r^{\mbox{\scriptsize mod}}(\lambda) =
    r(\lambda)\frac{\alpha(\lambda,\lambda_1)}{\alpha(\lambda_1,\lambda)}.
  \end{equation}
\end{lemma}
\begin{proof}
  $\M_n$ is reduced to $\M_{n-1}$
  with the help of commutation relation (\ref{cb})
  and relations (\ref{ay})--(\ref{ddy}).
  Letting both $A_{l+N-n+2,l-N+n}$ and $D_{l+N-n+1,l-N+n-1}$
  act to the right Bethe vector we obtain
  \begin{eqnarray}
    \label{nto}
    \lefteqn{\M_n(\lambda_1,\ldots,\lambda_n)}
    \nonumber \\ &=&
    \frac{1}{Q}\sum_{l=0}^{Q-1}\lim_{\lambda_1^C\rightarrow\lambda_1}
    [\beta_{l+N-n+1}(\lambda_1^C,\lambda_1)
    {}_1\Lambda(\lambda_1^C;\{\lambda_k\}_{k\ne 1})
    {}_2\Lambda(\lambda_1;\{\lambda_k\}_{k\ne 1})
    \nonumber \\ && \qquad\qquad
    -\beta_{l-N+n-1}(\lambda_1^C,\lambda_1)
    {}_1\Lambda(\lambda_1;\{\lambda_k\}_{k\ne 1})
    {}_2\Lambda(\lambda_1^C;\{\lambda_k\}_{k\ne 1})]
    \nonumber \\ && \qquad
    \times\langle\Psi_l^{n-1}(\lambda_2,\ldots,\lambda_n)|
    \Psi_l^{n-1}(\lambda_2,\ldots,\lambda_n)\rangle
    \nonumber \\ &&
    + \mbox{ terms independent of }X_1
    \nonumber \\ &=&
    h(2\eta)(h(\lambda_1+\eta)h(\lambda_1-\eta))^L
    \prod_{j\ne k}^n\alpha(\lambda_j,\lambda_k)
    \nonumber \\ &&
    \times \frac{1}{-h'(0)}
    \frac{\der}{\der\lambda_1}
    \log\left(
      r(\lambda_1)\prod_{k=2}^n
      \frac{\alpha(\lambda_1,\lambda_k)}{\alpha(\lambda_k,\lambda_1)}\right)
    \times\M_{n-1}^{\mbox{\scriptsize mod}}(\lambda_2,\ldots,\lambda_n)
    \nonumber \\ &&
    + \mbox{ terms independent of }X_1.
  \end{eqnarray}
  Here we have used the l'Hospital's rule.
  Notice that extra terms whose right Bethe vectors still contain $\lambda_1$
  do not generate $X_1$,
  because it raises only in the case that
  both of the Bethe vectors depend on $\lambda_1$
  and the l'Hospital's rule is applied.
  The formula (\ref{nto}) implies the lemma.
\end{proof}
\begin{lemma}
  \label{l4}
  $\|\lambda_1\|_1=X_1$.
\end{lemma}
\begin{proof}
  The proof is straightforward with $\tau_{l+N}=\tau_{l-N}$.
\end{proof}

By lemmas~\ref{l1}--\ref{l4}
$\|\lambda_1,\ldots,\lambda_n\|_n$ is determined uniquely.
The following is a main result of the paper
and corresponds to the Gaudin hypothesis for the $XYZ$ spin chain.
\begin{theorem}
  $\|\lambda_1,\ldots,\lambda_n\|_n$ has
  the following Jacobian form:
  \begin{equation}
    \label{result}
    \|\lambda_1,\ldots,\lambda_n\|_n =
    \mbox{$\det_n$}\frac{\der\phi_k}{\der\lambda_j},
  \end{equation}
  where
  \begin{equation}
    \phi_k =
    \log\left(r(\lambda_k)\prod_{i\ne k}^n
      \frac{\alpha(\lambda_k,\lambda_i)}{\alpha(\lambda_i,\lambda_k)}\right).
  \end{equation}
\end{theorem}
\begin{proof}
  It is obvious that this expression satisfies lemma~\ref{l1}--\ref{l4}.
  We prove its converse by induction on $n$.
  Let
  \begin{equation}
    \Delta_q =
    \|\lambda_1,\ldots,\lambda_q\|_q
    -\mbox{$\det_q$}\frac{\der\phi_k}{\der\lambda_j}.
    \qquad
    (q=1,\ldots,n)
  \end{equation}
  By lemma~\ref{l4} it follows that $\Delta_1=0$.
  Let us assume that $\Delta_q=0$ for $q=1,\ldots,n-1$.
  By lemma~\ref{l3} we have
  \begin{equation}
    \frac{\der\Delta_n}{\der X_1} =
    \|\lambda_2,\ldots,\lambda_n\|_{n-1}^{\mbox{\scriptsize mod}}
    -\mbox{$\det_{n-1}$}\frac{\der\phi_k^{\mbox{\scriptsize mod}}}
    {\der\lambda_j}.
  \end{equation}
  By assumption of induction
  the right hand side is equal to $0$.
  Thus $\Delta_n$ is independent of $X_1$.
  By lemma~\ref{l1} $\Delta_n$ does not depend on any $X_j$
  ($j=1,\ldots,n$).
  Hence we obtain $\Delta_n=0$ owing to lemma~\ref{l2}.
  The proof has been completed.
\end{proof}

The function $\phi_k$ is expanded as
\begin{eqnarray}
  \phi_k &=&
  2\pi\rmi l_k
  -L\left[\pi\rmi\left(1+\frac{\lambda_k}{K}\right)
    -2\sum_{m=1}^\infty
    \frac{\sin\frac{m\pi}{K}\lambda_k
      \sin\frac{m\pi}{K}(\eta-\frac{\rmii K'}{2})}
    {m\sinh\frac{m\pi K'}{2K}}\right]
  \nonumber \\ &&
  -\sum_{i\ne k}^n
  \left[\pi\rmi\left(1+\frac{\lambda_i-\lambda_k}{K}\right)
    -2\sum_{m=1}^\infty
    \frac{\sin\frac{m\pi}{K}(\lambda_i-\lambda_k)
      \sin\frac{m\pi}{K}(2\eta-\frac{\rmii K'}{2})}
    {m\sinh\frac{m\pi K'}{2K}}\right],
\end{eqnarray}
where $l_k$ is half-integer.
Because of the condition for $\eta$ (\ref{qm})
this series converge absolutely provided that
\begin{equation}
  \mbox{Im} \frac{\lambda_k}{K} = 0.
  \qquad
  (k=1,\ldots,n)
\end{equation}

\section{Concluding Remarks}
\label{con}

We have computed the sum of norms of the Bethe vectors
and have proved that it is expressed in the form of a Jacobian (\ref{result}).
Note that the Bethe vectors correspond to the eigenvectors
in the usual algebraic Bethe ansatz.
Our result is thus equivalent to
the Gaudin hypothesis for the $XYZ$ spin chain.

Physically, calculation of norms of the eigenvectors is important.
However, it is impossible to compute them
in the framework of the original generalized algebraic Bethe ansatz,
because extra scalar products of the Bethe vectors with different $l$
such that $\langle\Psi_l|\Psi_{l'}\rangle$ ($l\ne l'$) always appear,
and they can not be calculated.
It is necessary to develop a new method
to obtain not only norms of the eigenvectors
but also scalar products of arbitrary vectors
for the $XYZ$ spin chain.

\section*{Acknowledgment}
The authors thank to T. Deguchi, H. Fan and T. Tsuchida
for useful comments.

\end{document}